\documentclass[10pt]{article}

\usepackage[margin=1in]{geometry}
\usepackage{graphicx}
\usepackage{xcolor}
\usepackage{amsmath,amsfonts,amssymb}
\usepackage{booktabs}
\usepackage{nicefrac}
\usepackage{kotex}
\usepackage{subcaption}
\usepackage{float}
\usepackage{dblfloatfix}
\usepackage[ruled]{algorithm2e}
\usepackage[authoryear,square]{natbib}
\usepackage{hyperref}
\usepackage{authblk}

\SetAlFnt{\small}
\SetAlCapFnt{\small}
\SetAlCapNameFnt{\small}
\SetAlCapHSkip{0pt}
\IncMargin{-\parindent}

\title{Snapshot Polarimetric Display Inverse Rendering}

\author[1,*]{Seokjun Choi}
\author[1,*]{Yunseong Moon}
\author[1]{Kaizhang Kang}
\author[1]{Hoon-Gyu Chung}
\author[1]{Jin-Nyeong Kim}
\author[2]{Giljoo Nam}
\author[1]{Seung-Hwan Baek}

\affil[1]{POSTECH, South Korea}
\affil[2]{Meta, USA}
\affil[*]{Equal contribution.}

\date{}

\begin{document}

\maketitle

\begin{center}
\vspace{-8mm}
  \includegraphics[width=0.82\textwidth]{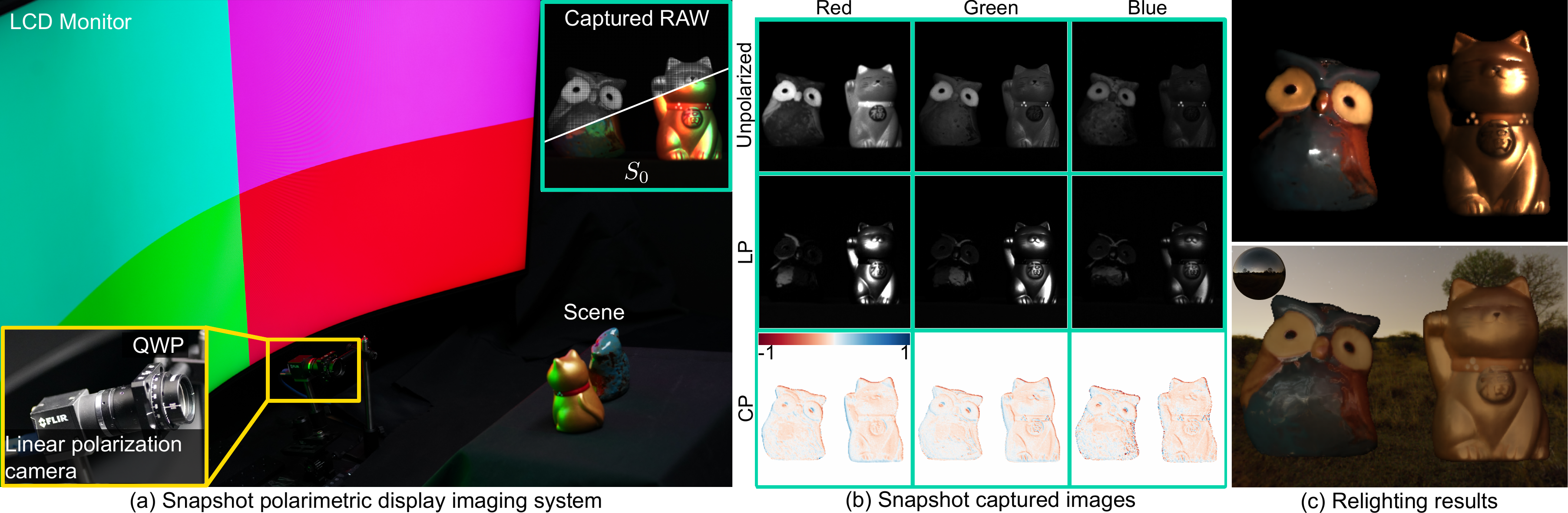}
  
  \begin{minipage}{0.82\textwidth}
  \footnotesize

  We present a snapshot feed-forward inverse rendering method that leverages spectro-polarimetric reflectance cues. 
A simple display-camera setup (a) captures a single-shot image, which (b) is decomposed into nine measurements encoding three RGB lighting directions and three polarization states: unpolarized, linearly-polarized (LP), circularly-polarized (CP). 
From these measurements, our network (c) predicts PBR maps that enable robust relighting under diverse illumination conditions.
  \end{minipage}
\end{center}

\begin{abstract}
Inverse rendering remains a core challenge in graphics and vision, especially in the snapshot configurations required for lightweight desktop workflows, where the per-frame information budget is highly constrained. Previous inverse rendering work explores various available dimensions for enriching the per-shot information, including temporal modulation, spectral encoding, and polarization. In this work, we introduce polarimetric display inverse rendering, using an LCD to project a linearly polarized RGB binary pattern and an RGB polarization camera augmented with a quarter-wave plate to acquire spectro-polarimetric measurements in a single shot. A feed-forward transformer maps these measurements to per-pixel normal, albedo, roughness, and metallicity. To overcome training data scarcity, we expand a limited set of measured polarimetric bidirectional reflectance distribution functions via a generative manifold. Evaluations on a real desktop setup demonstrate accurate inverse rendering across diverse scenes, outperforming existing approaches.
\end{abstract}


\newcommand{\vect}[1]{\mathbf{#1}}
\newcommand{\mat}[1]{\mathbf{#1}}

\newcommand\note[1]{\textcolor{red}{#1}}

\newcommand{\psf}{\rho}

\newcommand{\argmin}[1]{\stackrel[\{ #1 \}]{}{\textrm{arg min}}}
\newcommand{\minimize}[1]{\stackrel[\{ #1 \}]{}{\textrm{minimize}}}

\section{Introduction}
\label{sec:intro}
Inverse rendering--the estimation of geometry and {appearance} from images--has long stood as a foundational challenge in computer graphics and vision. {While decades of research, accelerated by recent learning-based solutions~\cite{li2025lirm, he2024neural, liang2025diffusion}, have produced high-quality results from multi-frame inputs, the snapshot setting required by lightweight desktop workflows remains challenging.}

Prior work exploits different dimensions of light to increase the capability of a single shot: high-speed temporal modulation that captures multiple lighting conditions within one exposure~\cite{wenger2005performance, sun2020light}, encoding multiple lighting directions across spectral channels~\cite{lattas2022practical, kampouris2018diffuse, ikehata2024physics}, and linear polarization that separates diffuse from specular reflectance~\cite{lattas2022practical, kampouris2018diffuse, choi2025real, ichikawa2024spiders}. 
This line of work shows that different light properties can provide additional information, motivating us to integrate spectral and polarimetric cues within one capture on a desktop setup.

In this work, we propose polarimetric display inverse rendering, a snapshot pipeline built around three coupled components: a display-camera imaging system for spectro-polarimetric acquisition, a feed-forward inverse rendering network, and a measured-reflectance dataset expansion method for training.

First, we construct a display-camera imaging system to capture a scene into nine spectro-polarimetric measurements within a single snapshot. On the illumination side, we use a liquid crystal display (LCD) to project a linearly polarized binary RGB pattern~\cite{lattas2022practical, kampouris2018diffuse, ikehata2024physics}, probing the scene simultaneously from multiple lighting directions across the R, G, and B channels. We then use an RGB polarization camera, which by itself records only the linear polarization components of the reflected light. We augment it with a quarter-wave plate (QWP), enabling single-shot acquisition of not only vertical linear polarization but also circular polarization, which provides cues for distinguishing dielectric and metallic reflectance.
The resulting observation forms a 3$\times$3 spectro-polarimetric representation: each of the three spectral channels is decomposed into three polarization states: unpolarized, vertically linearly polarized, and circularly polarized, which are related to diffuse, specular, and metallic reflectance components, respectively.

Second, we use a feed-forward network to estimate material parameters from these nine measurements. We modify a transformer-based architecture~\cite{li2025light} to treat the spectro-polarimetric measurements as a token sequence and exploit self-attention to integrate cues across polarization states and spectral channels, outputting per-pixel normal, albedo, roughness, and metallicity in a single forward pass.

Third, we develop a method that expands a limited set of measured polarimetric bidirectional reflectance distribution functions (pBRDFs)~\cite{baek2020image, moon2025hyperspectral} into a diverse, physically valid pBRDF dataset for training our neural network. 
Specifically, the measured pBRDFs are represented in a compact basis using principal component analysis (PCA), and a weight generator learns a generative manifold over the resulting PCA coefficients. New pBRDFs are synthesized by sampling weights from this manifold and reconstructing them through the PCA basis.
The synthesized pBRDFs are then assigned to geometry assets, enabling synthetic dataset rendering for training and testing the inverse rendering network.

By integrating these components, we demonstrate that our method enables accurate, single-shot inverse rendering on a desktop setup, outperforming baselines. Our main contributions are summarized as follows:

\begin{itemize}
\item A display-camera imaging system that probes the scene from multiple lighting directions across spectral channels with a linearly polarized RGB pattern and encodes linear-circular polarization in a single shot, producing nine spectro-polarimetric measurements.
\item A feed-forward transformer that estimates per-pixel normal, albedo, roughness, and metallicity from these spectro-polarimetric measurements in a single forward pass.
\item {A data-driven approach that expands a limited set of measured pBRDFs into a diverse and physically valid pBRDF dataset by sampling on the learned manifold.}
\end{itemize}

\section{Related Work}
\label{sec:related}
\paragraph{Inverse Rendering from Conventional Images}
Existing inverse-rendering methods often rely on multi-view~\cite{zhang2021nerfactor, zhang2021physg, zhang2022modeling, jin2023tensoir}, multi-light~\cite{li2022neural}, or both~\cite{zhang2022iron, yang2022ps, chung2024differentiable, chung2025differentiable} types of measurements.
Such settings have been widely adopted, particularly in NeRF-~\cite{mildenhall2021nerf} and 3DGS-based~\cite{kerbl20233d} methods, where iterative optimization has led to {high-quality} reconstruction performance.
However, these approaches typically require repeated captures of static scenes, making them less suitable for limited-observation settings.
In parallel, recent works have explored directly estimating scene properties from images using learned priors.
Diffusion-based methods~\cite{lyu2023diffusion, he2024neural, chen2025intrinsicanything, chen2025uni, liang2025diffusion} generate plausible results through iterative denoising guided by generative priors learned from large-scale data, requiring relatively fewer constraints on the capture setup.
However, they often fail to provide sufficient constraints for a physically valid disentanglement of geometry and reflectance.
Rather than relying on passive captures, our method encodes spectro-polarimetric information at capture time, providing a rich physical signal that complements learned priors.

\paragraph{Inverse Rendering by Spectro-polarimetric Imaging.}
Various active--illumination systems have been developed to provide informative observations for inverse rendering, including mobile flash cameras~\cite{lichy2021shape}, large-scale light stages~\cite{ghosh2009estimating, wenger2005performance, sun2020light}, and display--camera systems~\cite{aittala2013practical, choi2024differentiable, choi2025real, lattas2022practical}. 
{Display-based approaches provide spatially programmable area illumination,
enabling rich lighting cues to be multiplexed into fewer captures, but they still require multiple captures.}

Spectral multiplexing has been widely used to encode diverse lighting directions into a small number of measurements, using gradient~\cite{ma2007rapid, ghosh2009estimating, fyffe2015single}, binary~\cite{kampouris2018diffuse, lattas2022practical}, sinusoidal~\cite{lin2025single}, or learned illumination patterns~\cite{zhang2023deep, zhang2025sparse, choi2024differentiable, choi2025real, kang2019learning, ma2021free, kang2021neural}. 

Polarimetric imaging provides complementary cues because the polarization state of reflected light depends on geometry and reflectance, and has been used for shape-from-polarization~\cite{lyu2024sfpuel, chen2024pisr, deschaintre2021deep}, diffuse--specular separation~\cite{kampouris2018diffuse}, inverse rendering~\cite{li2024neisf, li2025neisf++, baek2018simultaneous, ichikawa2024spiders}, and spectro-polarimetric BRDF analysis~\cite{baek2020image, moon2025hyperspectral}. 
In particular, circular polarization has been shown to exhibit material-dependent behavior related to metallicity~\cite{baek2020image} and has been used in multi-frame reflectometry setups~\cite{ghosh2010circularly}, but has rarely been incorporated into snapshot inverse-rendering systems.

Our display--camera system combines spectral lighting multiplexing and linear--circular polarization decoding in a single shot. 
A linearly polarized RGB display pattern and a QWP before the RGB polarization camera enable richer spectro-polarimetric measurements for snapshot inverse rendering.

\paragraph{Data-driven Analysis and Expansion of BRDFs}
{Data-driven approaches have been used to model and expand BRDFs through PCA~\cite{nielsen2015optimal, sun2018connecting} and learned nonlinear manifolds~\cite{soler2018versatile}.
These ideas have been extended to pBRDFs through PCA-based analysis~\cite{baek2021polarimetric, moon2025hyperspectral}, but existing pBRDF studies focus on dimensionality reduction rather than generative sampling.}
{In this work, we first follow previous work by extracting principal components from measured pBRDFs. We then train a weight generator on PCA coefficients to learn a generative manifold, from which we sample new physically valid pBRDFs for large-scale training.}

\section{Overview}
\label{sec:overview}
A display--camera system encodes RGB lighting directions and polarization states into a single RAW capture, which is decomposed into nine spectro-polarimetric measurements (Sec.~\ref{sec:imaging}). 
A feed-forward transformer maps these measurements to spatially varying PBR maps, including normals, albedo, roughness, and metallicity (Sec.~\ref{sec:method}). 
For training, we combine real captures with synthetic data rendered from an expanded pBRDF dataset (Sec.~\ref{sec:dataset}).

\section{Imaging System}
\label{sec:imaging}
\subsection{Display-Camera Setup}
\begin{figure}[!t]
    \centering
    \includegraphics[width=0.55\linewidth]{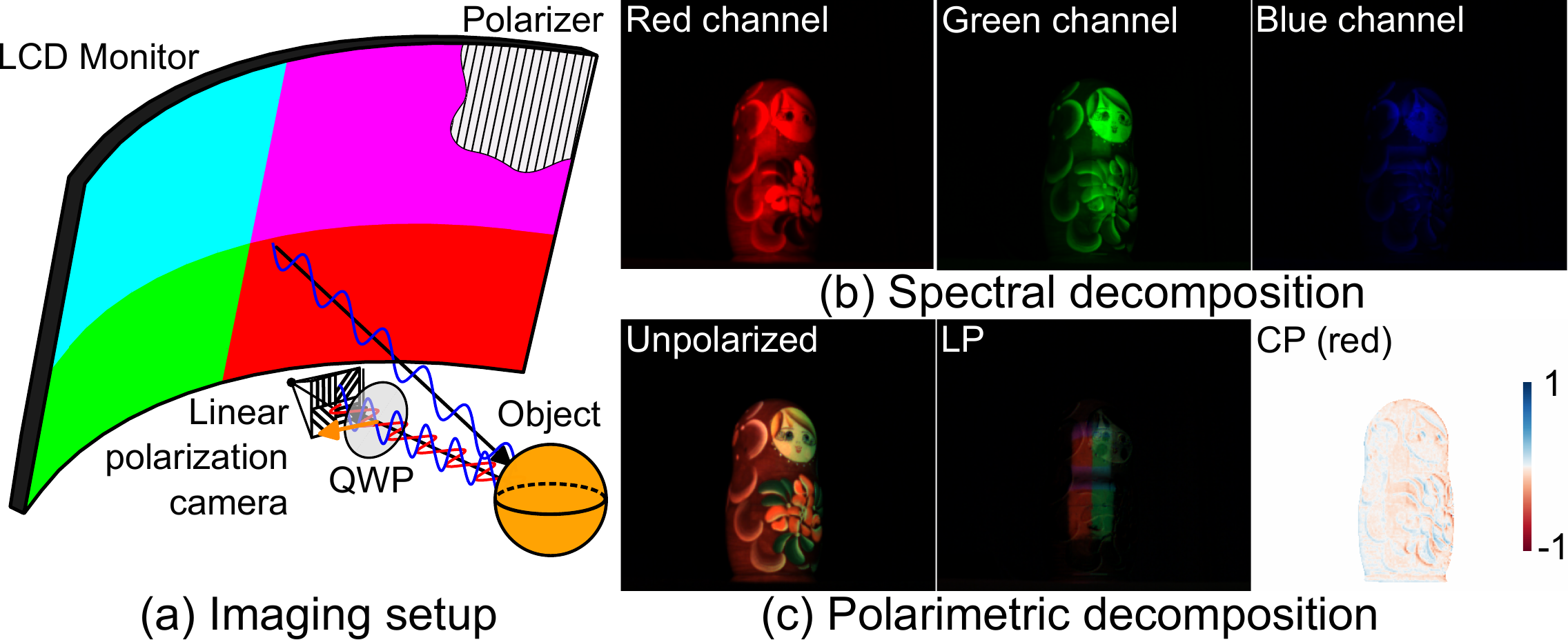}
    \caption{\textbf{Display–camera imaging system.} (a) Polarizers are mounted on both the illumination and the sensor, enabling polarimetric imaging. (b) The illumination pattern encodes distinct shading cues across the RGB channels. (c) A captured image decomposed into unpolarized, linearly polarized (LP), circularly polarized (CP) components.
    }
    \label{fig:system}
\end{figure}

We develop a display--camera imaging system for single-shot polarimetric acquisition, as shown in Fig.~\ref{fig:system}.
It consists of a 4K LCD display (Samsung Odyssey Ark, $600~\mathrm{cd/m^2}$ peak
luminance) and an RGB polarization camera (FLIR BFS-U3-51S5PC-C) equipped with an
$8\,\mathrm{mm}$ focal-length lens.
Due to the operating principle of liquid crystals~\cite{heilmeier1968guest}, the LCD emits
linearly polarized light, enabling it to serve simultaneously as a programmable spatial
light source and a polarized illuminator, with no additional polarizing optics on the
illumination side.

The reflected light at each pixel is described by the full Stokes vector~\cite{collett2005field}: $\mathbf{S}=\begin{bmatrix}
    S_0 & S_1 & S_2 & S_3
\end{bmatrix}^\mathsf{T}$, 
where $S_0$ denotes total intensity, $S_1$ and $S_2$ encode the difference of linearly polarized light between 0$^\circ$/90$^\circ$ degrees and 45$^\circ$/-45$^\circ$ respectively, and $S_3$ encodes the difference between right- and left-circular polarization.

Since a polarization camera with an on-sensor micro-polarizer array can only recover the linear Stokes components of the incident light, we place a QWP (Thorlabs AQWP10M-580) in front of the camera lens to additionally access circular polarization cues. Under the chosen axis alignment, the QWP modulates the reflected Stokes vector as
\begin{equation}
\mathbf{S}^{\mathrm{mod}}=
\begin{bmatrix}
    S_0^\mathrm{mod} \\
    S_1^\mathrm{mod} \\
    S_2^\mathrm{mod} \\
    S_3^\mathrm{mod} \\
\end{bmatrix}
=
\underbrace{
\begin{bmatrix}
    1 & 0 & 0 & 0\\
    0 & 1 & 0 & 0\\
    0 & 0 & 0 & 1\\
    0 & 0 & -1 & 0
\end{bmatrix}
}_{\mathbf{M}_{\mathrm{QWP}}}
\begin{bmatrix}
    S_0 \\
    S_1 \\
    S_2 \\
    S_3 \\
\end{bmatrix}
=\begin{bmatrix}
    S_0 \\
    S_1 \\
    S_3 \\
    -S_2
\end{bmatrix}.
\label{eq:qwp_components}
\end{equation}
The polarization camera then records the modulated light through its four on-sensor polarization orientations ($0^\circ$, $45^\circ$, $90^\circ$, $135^\circ$). Let $I_0$, $I_{45}$, $I_{90}$, and $I_{135}$ denote the corresponding measured intensities. From these measurements, we recover
\begin{equation}
\begin{aligned}
S_0^{\mathrm{mod}} &= \tfrac{1}{2}(I_0 + I_{45} + I_{90} + I_{135}),\\
S_1^{\mathrm{mod}} &= I_{0} - I_{90},\\
S_2^{\mathrm{mod}} &= I_{45} - I_{135},
\end{aligned}
\label{eq:stokes_linear}
\end{equation}
which correspond to $S_0$, $S_1$, and $S_3$ of the reflected light, respectively (Eq.~\eqref{eq:qwp_components}).

\subsection{Data Acquisition and Preprocessing}
\label{sec:acquisition}
\paragraph{Capture.}
{ We use the RGB binary illumination pattern~\cite{lattas2022practical}, shown in Fig.~\ref{fig:system}: the right half of the monitor (\(+X\)) is assigned to the red channel \(R\), the left half (\(-X\)) to the green channel \(G\), and the upper half (\(+Y\)) to the blue channel \(B\), resulting in a four-color pattern composed of cyan, magenta, green, and red regions.
This spectrally multiplexed design integrates three directional lighting cues into a single shot for shape and reflectance estimation.}
We {use the aforementioned polarization camera to capture raw measurements and} then compute the Stokes vector components {using} Eq.~\eqref{eq:stokes_linear}.

\paragraph{Spectral-polarimetric decomposition.}
Under vertically polarized LCD illumination, the component $S_1$ {captures the linearly polarized component (LP) of the reflected light, which is dominated by polarization-preserving specular reflection. Subtracting $S_1$ from the total intensity $S_0$ yields a diffuse-dominant unpolarized image (unpol). Finally, we recover the circularly polarized component (CP) from the normalized $S_3$ signal.} For each spectral channel $c \in \{R,G,B\}$, these three quantities are given by

\begin{equation}
I_{c,\mathrm{unpol}} = S_{0,c} - S_{1,c}, \qquad
I_{c,\mathrm{LP}}   = -S_{1,c}, \qquad
I_{c,\mathrm{CP}}   = \frac{S_{3,c}}{S_{0,c} + \epsilon},
\label{eq:decomp}
\end{equation}
where $\epsilon$ is a small constant for numerical stability. 
This yields nine per-pixel cues in total, which are collected as
\begin{equation}
\mathcal{I} =
\left\{
I_{R,\mathrm{unpol}}, I_{G,\mathrm{unpol}}, I_{B,\mathrm{unpol}},
I_{R,\mathrm{LP}}, I_{G,\mathrm{LP}}, I_{B,\mathrm{LP}},
I_{R,\mathrm{CP}}, I_{G,\mathrm{CP}}, I_{B,\mathrm{CP}}
\right\}.
\label{eq:input_9_images}
\end{equation}

All quantities above are defined at the pixel level; we compute them independently for each pixel and assemble the results into image maps, which constitute the network input. In the following, we use $\mathcal{I}=\{I_n\}_{n=1}^{9}$ to denote the resulting set of nine measurement maps (images), and $I_n(x)$ to denote the value of map $I_n$ at pixel $x$.

\section{Feed-forward Inverse Rendering}
\label{sec:method}
\begin{figure*}[t]
    \includegraphics[width=\linewidth]{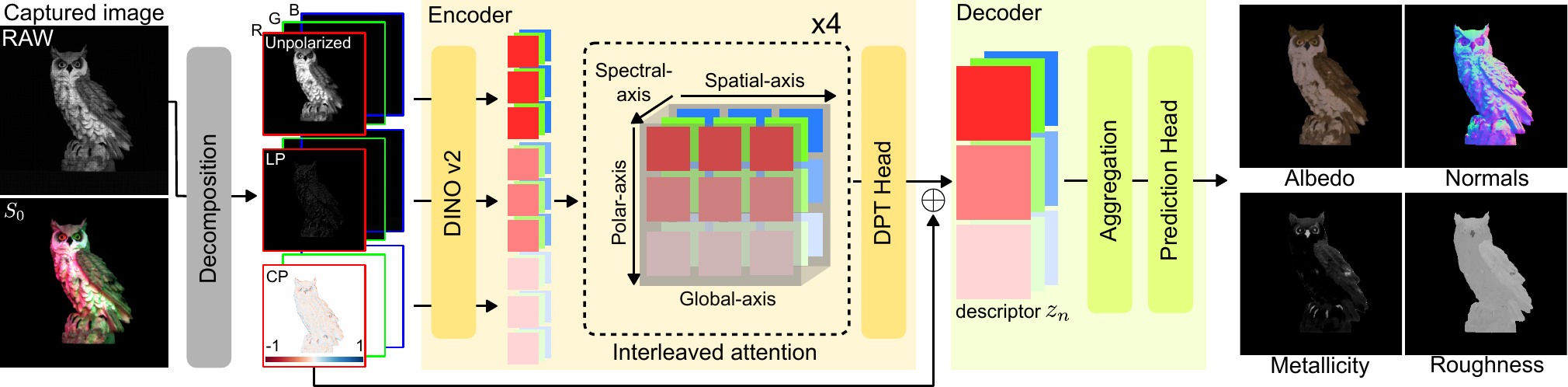}
    \caption{
    \textbf{Overview of feed-forward inverse rendering.}
    Our framework takes a single spectro-polarimetrically encoded RAW image as input and decomposes it into nine measurements spanning RGB lighting directions and polarization states. 
    An encoder--decoder transformer then estimates PBR parameter maps in a feed-forward manner.
    }
    \label{fig:overview}
\end{figure*}

Fig.~\ref{fig:overview} summarizes our feed-forward inverse rendering framework.
Given the nine decomposed polarimetric measurement maps $\mathcal{I}$ (Eq.~\eqref{eq:input_9_images}), our goal is to predict spatially-varying Disney~\cite{burley2012physically} physically-based rendering (PBR) parameter maps $\mathcal{G}=\{\mathbf{n},\mathbf{k},r,m\}$, where $\mathbf{n}$ denotes surface normals, $\mathbf{k}$ denotes albedo, $r$ denotes roughness, and $m$ denotes metallicity. Below we first describe the network architecture, followed by the training procedure using a hybrid dataset. 

\subsection{Network Architecture}
\label{sec:architecture}

Our network adopts an encoder--decoder architecture that aggregates information across the spatial, spectral, and polarization dimensions of the decomposed measurements, which is inspired by recent inverse rendering frameworks~\cite{ikehata2023scalable, li2025light}. 

\paragraph{Encoder.} The encoder processes each input measurement map in $\mathcal{I}$ (Eq.~\eqref{eq:input_9_images}) by a ViT-based feature extractor~\cite{oquab2023dinov2}, which converts it into a sequence of patch tokens. We then aggregate information both within each token sequence and across sequences to exploit the spatial, spectral, and polarimetric structure of the inputs. Specifically, the encoder applies \(K{=}4\) iterations of interleaved self-attention. Each iteration consists of four sequential steps: spatial attention within each observation, spectral attention across the three spectral channels sharing the same polarimetric modality, cross-observation attention across all nine observations, and polarimetric attention across the three polarimetric modalities sharing the same spectral channel. After the attention iterations, each patch token integrates context from corresponding positions across all nine decomposed measurements. A dense prediction transformer (DPT) module~\cite{ranftl2021vision} then decodes each enriched token sequence into a feature map \(F_{\mathrm{enc},n} \in \mathbb{R}^{H' \times W' \times D}\), where \(H' \times W'\) is a predefined canonical spatial resolution and \(D\) is the feature dimension.

\paragraph{Decoder.} The decoder takes the encoded features from all inputs and produces per-pixel geometry and reflectance descriptors for final prediction. For each pixel location \(x\), it first constructs a per-input descriptor by combining the encoded feature with the corresponding decomposed measurement:
\begin{equation}
z_n(x) = \phi\!\big(F_{\mathrm{enc},n}(x)\big) + \psi\!\big(I_n(x)\big),
\end{equation}
where \(\phi(\cdot)\) and \(\psi(\cdot)\) are linear projection layers. 
This fusion combines both high-level material cues from the encoder and local appearance evidence from the input. 
The descriptors \(\{z_n(x)\}_{n=1}^{9}\) are then aggregated through a cross-attention module with a learnable query vector, producing a single fused descriptor \(a(x) \in \mathbb{R}^{D_a}\), where \(D_a\) denotes the feature dimension of the aggregated descriptor:
\begin{equation}
a(x) = \mathrm{Agg}\Big(\{z_n(x)\}_{n=1}^{9}\Big).
\end{equation}

For memory efficiency, the decoder performs this aggregation on a set of randomly sampled pixel locations during training. 
The sampled descriptors are further allowed to interact through a transformer module, enabling non-local information exchange before prediction. 

Finally, the network predicts the PBR parameters using per-pixel heads for geometry and reflectance:
\begin{equation}
\hat{\mathbf{n}}(x) = h_{\mathrm{normal}}\!\big(a(x)\big), \qquad
\big(\hat{\mathbf{k}}(x), \hat{r}(x), \hat{m}(x)\big) = h_{\mathrm{PBR}}\!\big(a(x)\big),
\end{equation}
where \(\hat{\mathbf{n}}(x)\), \(\hat{\mathbf{k}}(x)\), \(\hat{r}(x)\), and \(\hat{m}(x)\) denote the predicted surface normal, albedo, roughness, and metallic value, respectively. We use $\hat{\mathcal{G}}$ to denote the collection of those predicted maps. Additional implementation details are provided in the supplementary material.

\subsection{Training}
\label{sec:training}

We train the network on a hybrid dataset consisting of synthetic and real captures. Synthetic samples provide ground-truth parameter maps $\mathcal{G}_{\mathrm{GT}}$, enabling direct supervision of the network. In contrast, although real samples do not have ground-truth PBR parameters,  they help bridge the synthetic--real domain gap. 
During training, we use a joint data loader that provides one synthetic batch and one real batch at each iteration. 
The network is optimized with the weighted objective
\begin{equation}
L_{\mathrm{total}} = w_s L_{\mathrm{synth}} + w_r L_{\mathrm{real}},
\end{equation}
where \(w_s=1\) and \(w_r\) is gradually increased with a warm-up schedule. 
This schedule stabilizes early training under synthetic PBR supervision before gradually introducing real-domain reconstruction constraints.

\paragraph{Synthetic data.} We construct a large-scale dataset using Mitsuba3~\cite{jakob2022mitsuba3}. Scene geometry is randomly sampled from Objaverse~\cite{deitke2023objaverse} and placed at random poses within the replicated display-camera setup in simulation. We set that the simulated display emits vertically linearly polarized light with $\mathbf{s}_{\mathrm{LCD}} = [1,\,-1,\,0,\,0]^\top$. Each object is assigned a pBRDF from the expanded pBRDF dataset described in Sec.~\ref{sec:dataset}. Each sample provides ground-truth PBR maps $\mathcal{G}_{\mathrm{GT}}$ alongside the nine decomposed polarimetric maps $\mathcal{I}$, enabling direct parameter supervision:
{\begin{equation}
L_{\mathrm{synth}} =
\left\| \hat{\mathcal{G}} - \mathcal{G}_{\mathrm{GT}} \right\|_2^2
+ \lambda_{\mathrm{chr}} L_{\mathrm{chr}}.
\end{equation}
Here, $L_{\mathrm{chr}}$ denotes an albedo chromaticity loss that compares brightness-normalized albedo values, encouraging color recovery independently of intensity scale.}

\paragraph{Real data.}
Since real samples do not provide ground-truth PBR parameters, we supervise the network by re-rendering the six intensity-based measurements in \(\mathcal{I}\): the specular-sensitive \(I_{c,\mathrm{LP}}\) and diffuse-dominant \(I_{c,\mathrm{unpol}}\) components for \(c\in\{R,G,B\}\). 
Given the predicted maps \(\hat{\mathcal{G}}\), we render each component as
\begin{equation}
\hat{I}_{c,\mathrm{LP}}(x)
= \int_{\Omega} f_{s}\!\big(\omega_i,\omega_o;\hat{\mathcal{G}}(x)\big)\,
L(\omega_i,c)\,[\mathbf{n}(x)\!\cdot\!\omega_i]_+\,\mathrm{d}\omega_i,
\end{equation}
\begin{equation}
\hat{I}_{c,\mathrm{unpol}}(x)
= \int_{\Omega} f_{d}\!\big(\omega_i,\omega_o;\hat{\mathcal{G}}(x)\big)\,
L(\omega_i,c)\,[\mathbf{n}(x)\!\cdot\!\omega_i]_+\,\mathrm{d}\omega_i,
\end{equation}
where \(f_s\) and \(f_d\) denote the specular and diffuse BRDF components corresponding to \(I_{c,\mathrm{LP}}\) and \(I_{c,\mathrm{unpol}}\), respectively. 
The real-data reconstruction loss is
\begin{equation}
L_{\mathrm{real}} =
\sum_{c}
\left(
\left\| \hat{I}_{c,\mathrm{LP}} - I_{c,\mathrm{LP}} \right\|_2^2
+
\left\| \hat{I}_{c,\mathrm{unpol}} - I_{c,\mathrm{unpol}} \right\|_2^2
\right).
\end{equation}

\section{Expanded Polarimetric BRDF Dataset}
\label{sec:dataset}
To generate diverse and realistic synthetic training samples, we need a large collection of physically valid polarimetric BRDFs. While the measured pBRDF datasets~\cite{baek2020image, moon2025hyperspectral} provide high-fidelity reflectance measurements, they contain only a limited number of materials (39 in total). We therefore construct an expanded pBRDF dataset. 

\begin{figure}[t]
    \centering
    \includegraphics[width=0.55\linewidth]{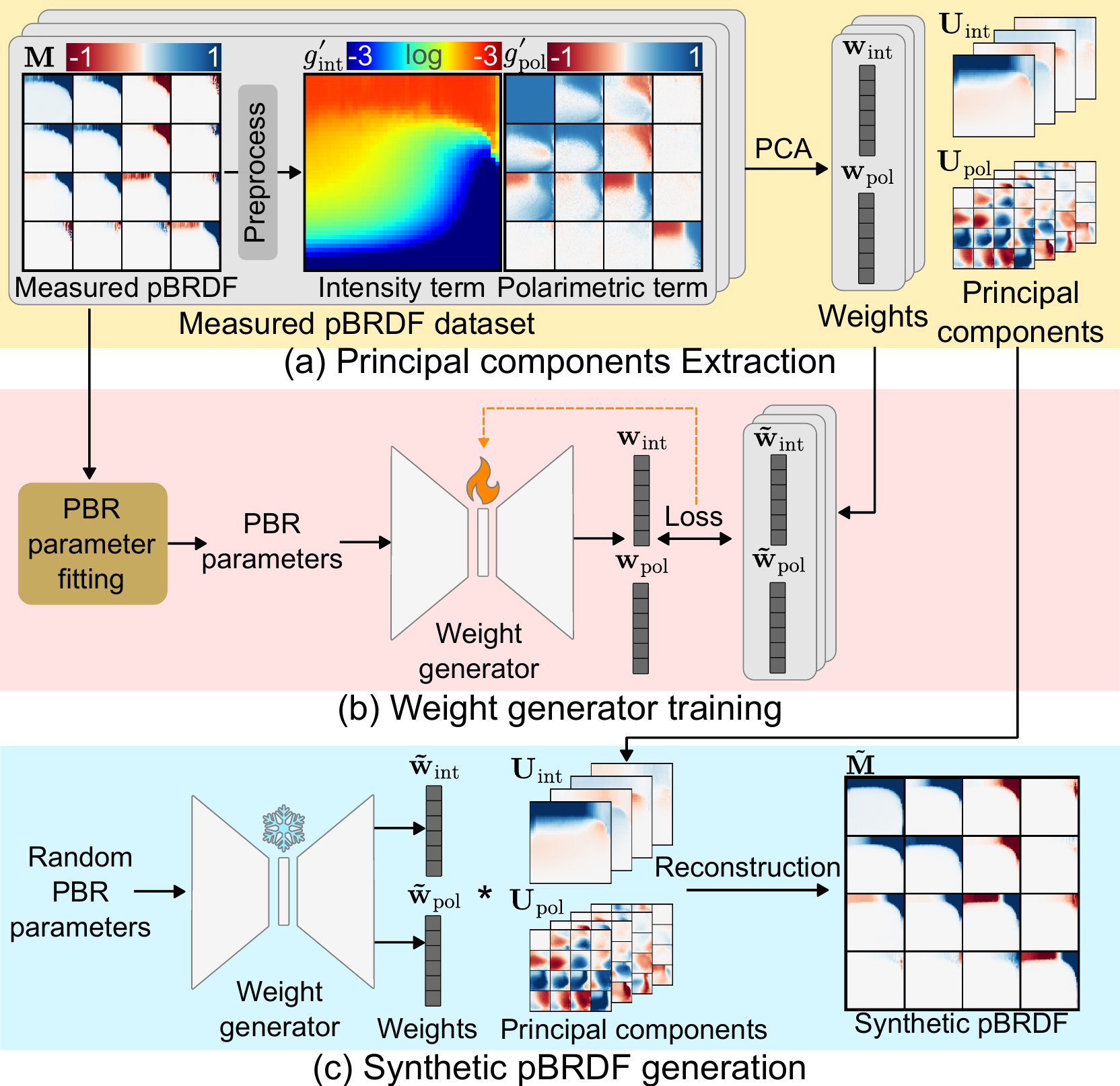}
    \caption{
    \textbf{Overview of the expanded pBRDF dataset generation pipeline.}
    (a) Principal components and corresponding weights are extracted from measured pBRDF dataset via PCA, separately for intensity and polarimetric components.
    (b) A weight generator is trained to predict PCA weights conditioned on input PBR parameters.
    (c) After training, the generator samples weights from randomly sampled PBR parameters, and synthetic pBRDFs are reconstructed using the precomputed principal components from (a).
    }
    \label{fig:brdf_expand}
\end{figure}

The expanded pBRDF dataset is constructed through the three-stage pipeline shown in Fig.~\ref{fig:brdf_expand}: (a) measured pBRDFs are preprocessed and compressed into a compact low-dimensional representation via PCA, (b) the resulting PCA weight distribution is modeled by a weight generator conditioned on the PBR parameters, and (c) we generate new weights and reconstruct novel pBRDFs via PCA inversion. The reconstructed pBRDFs are subsequently filtered with a physical-validity check. 

\subsection{pBRDF Preprocessing}
As illustrated in Fig.~\ref{fig:brdf_expand}(a), we preprocess the previously measured pBRDF tables~\cite{baek2020image, moon2025hyperspectral} for PCA.
The pBRDF tables contain a Mueller matrix $\mathbf{M}$ for each incident/outgoing pair $(\omega_i,\omega_o)$ and wavelength. {
{Since the magnitude of \(\mathbf{M}\) varies significantly across materials and angles, we factor the pBRDF into an intensity component and a normalized polarimetric component as
\begin{equation}
\mathbf{M}(\omega_i,\omega_o)
= m_{00}(\omega_i,\omega_o)\,\mathbf{M}_{\mathrm{pol}}(\omega_i,\omega_o),
\label{eq:pbrdf_factor}
\end{equation}
where \(m_{00}\) is the \((0,0)\) entry of \(\mathbf{M}\), and
\(\mathbf{M}_{\mathrm{pol}}=\mathbf{M}/m_{00}\) is the normalized Mueller matrix that captures polarization behavior independent of scale.}
We stack all samples of $m_{00}(\omega_i,\omega_o)$ and $\mathbf{M}_\mathrm{pol}(\omega_i,\omega_o)$ over the sampled $(\omega_i,\omega_o)$ tuples and vectorize them into $\mathbf{g}_{\mathrm{int}}$ and $\mathbf{g}_{\mathrm{pol}}$, respectively.
To learn a PCA basis that is shared across wavelengths, we flatten the wavelength dimension, regarding each per-wavelength pBRDF as an independent sample.

To improve numerical stability of PCA, we follow \citet{baek2021polarimetric} by applying nonlinear compression and mean-centering. Specifically, for each term $t\in\{\mathrm{int},\mathrm{pol}\}$ with vectorized representation $\mathbf{g}_t$, we apply:
\begin{equation}
\mathbf{g}'_t = \arctan\!\bigl(\alpha \mathbf{g}_t\bigr) - \boldsymbol{\mu}_t,
\label{eq:preproc}
\end{equation}
where $\alpha$ is a scaling constant and $\boldsymbol{\mu}_t$ denotes the dataset mean of $\arctan\!\bigl(\alpha \mathbf{g}_t\bigr)$, computed element-wise.
The resulting mean-centered vectors $\mathbf{g}'_\mathrm{int}$ and $\mathbf{g}'_\mathrm{pol}$ are used as the inputs to PCA decomposition.

\subsection{PCA Decomposition}

After preprocessing, we perform PCA separately on the intensity and polarimetric terms, i.e., on $\mathbf{g}'_\mathrm{int}$ and $\mathbf{g}'_\mathrm{pol}$ shown in Fig.~\ref{fig:brdf_expand}(a).

For each term $t$, PCA yields a basis matrix $\mathbf{U}_t\in\mathbb{R}^{d_t\times p_t}$ whose columns are the top-$p_t$ principal components, where \(d_t\) is the dimensionality of the vectorized term and \(p_t\) is the number of principal components.
We retain $p_{\mathrm{int}}=p_{\mathrm{pol}}=10$ principal components for both terms, which together capture over 90\% of variance in the measured pBRDF set.
Given an instance $\mathbf{g}'_t\in\mathbb{R}^{d_t}$, we compute its PCA weights by projection:
\begin{equation}
\mathbf{w}_t = \mathbf{U}_t^{\mathsf T}\,{\mathbf{g}'}_t,
\label{eq:pca_project}
\end{equation}
and reconstruct the centered compressed representation as
\begin{equation}
\hat{\mathbf{g}'}_t = \mathbf{U}_t\mathbf{w}_t.
\label{eq:pca_recon_centered}
\end{equation}

\subsection{PCA Weight Sampling}
PCA provides a compact representation of each pBRDF via weights $\mathbf{w}_t\in\mathbb{R}^{p_t}$. The second stage of our pipeline learns a generative model that maps Disney PBR parameters $\{\mathbf{k},r,m\}$ to PCA weights $\mathbf{w}=(\mathbf{w}_\mathrm{int},\mathbf{w}_\mathrm{pol})$, which are decoded into a Mueller-matrix-valued pBRDF $\mathbf{M}(\omega_i,\omega_o)$ through PCA inversion and Eq.~\eqref{eq:pbrdf_factor}. We instantiate this generative model as a conditional variational autoencoder-based generator, shown in Fig.~\ref{fig:brdf_expand}(b), which models the conditional distribution
\begin{equation}
p(\mathbf{w} \mid \mathbf{k}, r, m)
\;=\; \int p_\psi(\mathbf{w} \mid \mathbf{z})\, q_\phi(\mathbf{z} \mid \mathbf{k}, r, m)\, d\mathbf{z},
\label{eq:cvae_def}
\end{equation}
where $p_\psi$, $q_\phi$ are the encoder, decoder and $\mathbf{z}$ is the latent vector.
A simple parametric distribution over $\mathbf{w}$ would not respect the constrained manifold of physically valid pBRDFs, while the empirical correlations between $\{\mathbf{k},r,m\}$ and $\mathbf{w}$ make the Disney PBR parameters an informative conditioning signal. The weight generator addresses both observations simultaneously. We provide the supporting empirical analysis along with the encoder/decoder architecture and training details in the Supplementary Document.

At inference time, illustrated in Fig.~\ref{fig:brdf_expand}(c), the trained weight generator generates new weights $\tilde{\mathbf{w}}_t$ by sampling in the latent space conditioned on input parameters. The sampled weights are mapped back by PCA reconstruction,
\begin{equation}
\tilde{\mathbf{g}'}_t = \mathbf{U}_t\tilde{\mathbf{w}}_t,
\end{equation}
followed by adding the dataset mean and applying the inverse of Eq.~\eqref{eq:preproc}.
Applying this procedure to both $t=\mathrm{int}$ and $t=\mathrm{pol}$ yields a synthesized intensity term $\tilde{m}_{00}$ and a synthesized normalized Mueller term $\tilde{\mathbf{M}}_{\mathrm{pol}}$. These are combined via Eq.~\eqref{eq:pbrdf_factor} to form a novel pBRDF for rendering synthetic training data.
Fig.~\ref{fig:pbrdf_expansion} visualizes representative synthesized pBRDFs, showing diverse appearance, polarimetric responses, and fitted PBR parameters used for synthetic supervision.

\begin{figure}[t]
\centering
\includegraphics[width=0.65\linewidth]{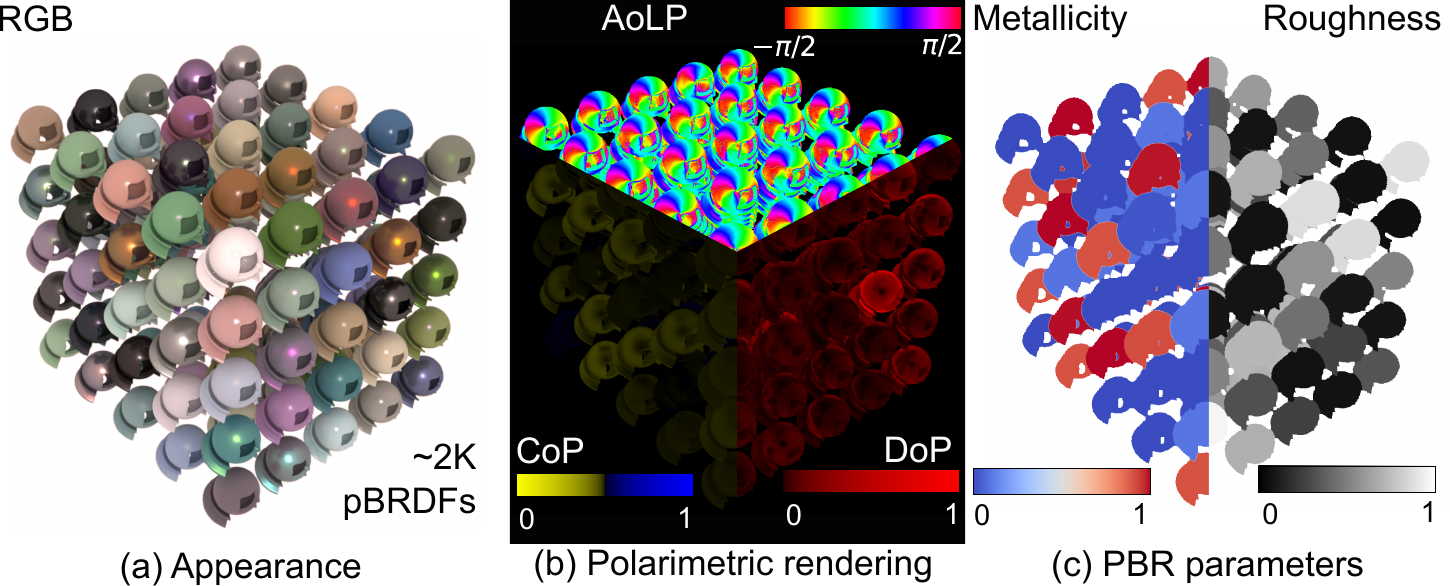}
\caption{\textbf{Expanded pBRDF instances.} Rendered examples of synthesized pBRDFs showing diverse reflectance.
(a) RGB appearance under unpolarized illumination.
(b) Polarimetric rendering visualized as {AoLP (Angle of Linear Polarization), DoP (Degree of Polarization), and CoP (Chirality of Polarization)}.
(c) PBR parameter maps fitted to each synthesized pBRDF and used as ground-truth supervision during training.
}
\label{fig:pbrdf_expansion}
\end{figure}

\section{Results}
\label{sec:results}
\subsection{Experimental Settings}

The spatial position and radiometric nonlinearity of the LCD and camera are calibrated following~\cite{choi2025real}. 
For the pBRDF representation, we retain the top-10 principal components for each intensity and polarization part. 
From the expanded samples, we collect 2{,}000 valid pBRDF instances after filtering based on physical validity constraints using Givens-Kostinski criterion~\cite{givens1993simple} and Cloude filtering~\cite{cloude1990conditions}.
For synthetic data generation, we screen 10{,}000 objects from Objaverse and randomly sample 10 objects as a subset. 
Each object is assigned a random pBRDF and placed within the camera field of view at a distance of 50--100 cm with random pose, yielding 12k scenes in total. 
For real data, we capture 373 scenes from 91 objects with diverse materials, including metal, rubber, plastic, wood, stone, resin, and plaster, using HDR exposure brackets of \(\{10,20,100,200\}\,\mathrm{ms}\) to reduce saturation.
All images are resized to \(384 \times 384\). 
The encoder produces feature maps at \(192 \times 192\) resolution with dimension \(D=256\), and the aggregated descriptor has dimension \(D_a=384\).
We implement our framework in PyTorch and train it with AdamW on four NVIDIA A6000 GPUs for 15{,}000 iterations with a learning rate of \(8\times10^{-5}\) and a per-GPU batch size of 1; training takes approximately one day.

\paragraph{Baselines.}
We compare against LINO-PBR (inverse rendering variant of LINO-UniPS~\cite{li2025light}), DiffusionRenderer~\cite{liang2025diffusion}, and RGB-X~\cite{zeng2024rgb}. 
Since these methods do not assume spectral multiplexing, we provide images rendered or captured under white illumination as input. 
DiffusionRenderer, which is designed for video inputs, is fed a static sequence constructed by replicating a single frame 24 times. 
LINO-PBR additionally supports multiple inputs, and we therefore evaluate a four-light setting (\(N=4\)) to assess whether multi-shot observations can compensate for the lack of multiplexed encoding.

\begin{table}[!t]
\centering
\caption{\textbf{Quantitative comparison on synthetic dataset.} We report PSNR and LPIPS (albedo), RMSE (roughness/metallicity), and angular error (normal). \textbf{Bold} indicates the best result, and \underline{underlined} text indicates the second-best result. (LINO-PBR ($N{=}4$)$^\dagger$ takes four-shot input, shown for reference.)}
\label{tab:quantitative}
\resizebox{\linewidth}{!}{
    \begin{tabular}{l|cccc|ccc}
    \toprule
     & \multicolumn{4}{c|}{Albedo}
     & Roughness
     & Metallicity
     & Normals \\
    \cmidrule(lr){2-8}
     Method
     & PSNR $\uparrow$ & LPIPS $\downarrow$ & si-PSNR $\uparrow$ & si-LPIPS $\downarrow$
     & RMSE $\downarrow$
     & RMSE $\downarrow$
     & MAE $\downarrow$ \\
    \midrule
    DiffusionRenderer    & 13.03 & 0.4039 & 17.61 & 0.3557 & 0.6999 & 0.6317 & 29.73 \\
    RGB-X                 & 4.75 & 0.4884 & \underline{18.89} & \underline{0.3199} & \underline{0.3596} & \underline{0.5529} & 40.47 \\
    LINO-PBR ($N=1$)      & 14.80 & 0.4446 & 16.56 & 0.4175 & 0.6658 & 0.6476 & 35.64 \\
    Ours                  & \textbf{23.90} & \textbf{0.1788} & \textbf{24.48} & \textbf{0.1736} & \textbf{0.0781} & \textbf{0.2395} & \textbf{11.45} \\
    \midrule
    LINO-PBR ($N=4$)$^\dagger$      & \underline{15.60} & \underline{0.3708} & 17.77 & 0.3422 & 0.3746 & 0.5843 & \underline{18.23} \\
    \bottomrule
    \end{tabular}
}
\end{table}

\subsection{Quantitative Evaluation}
Tabs.~\ref{tab:quantitative} and~\ref{tab:real_metrics} report quantitative comparisons on synthetic and real data. 
Scale-invariant metrics, denoted by ``si-'', are computed after aligning the prediction to the reference with a single global intensity scale over valid pixels.
On synthetic data, where ground-truth PBR maps are available, our method achieves the best performance across all metrics, with particularly clear gains in albedo and metallicity, indicating reduced ambiguity among illumination, reflectance, and material type. 
Although RGB-X produces saturated albedo and low standard PSNR, its scale-invariant albedo metric improves to the second-best result, suggesting that its main error lies in intensity scale rather than chromatic structure.

For real data, ground-truth PBR maps are unavailable; we therefore evaluate reconstruction error by re-rendering the estimated PBR maps under the captured pattern illumination. 
We report the average over 12 target patterns, which measures reconstruction consistency on the display setup rather than generalization to arbitrary incident illumination. 
Despite using a single-shot input, our method achieves competitive performance with the multi-light LINO-PBR variant (\(N=4\)), even though the input formats differ.

\begin{table}[!ht]
\centering
\caption{\textbf{Quantitative comparison on real data.} 
We report reconstruction error between the observation and the re-rendered image using the estimated PBR parameters. \textbf{Bold} indicates the best result, and \underline{underlined} text indicates the second-best result. (LINO-PBR ($N{=}4$)$^\dagger$ takes four-shot input, shown for reference.)}
\label{tab:real_metrics}
\resizebox{0.75\linewidth}{!}{
\begin{tabular}{l|cccc}
\toprule
Method & si-PSNR $\uparrow$ & si-LPIPS $\downarrow$ & si-RMSE $\downarrow$ & si-MSE $\downarrow$ \\
\midrule
DiffusionRenderer       & 16.31 & 0.2729 & 0.1652 & 0.0318 \\
RGB-X                   & 12.75 & 0.4121 & 0.2428 & 0.0651 \\
LINO-PBR ($N=1$)          & {19.16} & 0.1949 & \underline{0.1220} & 0.0186 \\
Ours                    & \textbf{20.15} & \underline{0.1690} & \textbf{0.1050} & \textbf{0.0126} \\
\midrule
LINO-PBR ($N=4$)$^\dagger$          & \underline{19.18} & \textbf{0.1493} & {0.1223} & \underline{0.0183} \\
\bottomrule
\end{tabular}
}
\end{table}

\subsection{Diffuse--Specular and Metallic Disambiguation}
We analyze the predicted albedo and metallicity maps in Figs.~\ref{fig:comparison}, ~\ref{fig:albedo}, and~\ref{fig:metallic}. 
Fig.~\ref{fig:comparison} compares the predicted PBR maps and the rendered images of our method against the baselines on a representative real scene, exposing two ambiguities that persist even when LINO-PBR is given $N{=}4$ multi-light inputs: diffuse--specular entanglement and dielectric--metallic confusion.
Fig.~\ref{fig:albedo} shows that single-image baselines often bake reflected illumination into albedo, leaving residual shading and highlight imprints, whereas our polarization decomposition assigns reflected energy to the appropriate component. 
Fig.~\ref{fig:metallic} examines the dielectric--metallic ambiguity through predicted metallicity.
{The CP input provides a physical cue for the dielectric--metallic ambiguity through the opposite circular-polarization handedness of visually similar samples.}
Baselines without CP input often fail to clearly separate metallic regions, producing visually plausible but physically inconsistent reflectance estimates. 
This leads to noticeable relighting errors, especially in highlight intensity and shape. 
In contrast, our method leverages CP cues as an additional constraint for metallic reflectance, resulting in more consistent metallicity estimates and lower re-rendering error.

\subsection{Ablation Study}

\paragraph{Effect of spectro-polarimetric decomposition.}
{Tab.~\ref{tab:ablation} evaluates the contribution of each input component. 
Spectral multiplexing substantially improves geometry estimation, reducing normal MAE from \(20.08\) to \(12.45\), by encoding distinct lighting directions across RGB channels. 
Adding polarization decomposition progressively reduces material ambiguity, improving albedo PSNR from \(21.61\) to \(22.22\) and roughness RMSE from \(0.1024\) to \(0.0889\). 
Finally, adding the CP cue yields the largest gain in metallicity, reducing metallicity RMSE from \(0.2889\) to \(0.2395\), indicating improved dielectric--metallic separation.}

\paragraph{Effect of expanded pBRDF dataset.}
We compare our measured-pBRDF expansion with analytically generated
pBRDFs over two held-out test splits. 
While the expanded dataset improves reflectance-related metrics, its main
benefit is reducing the domain gap between synthetic polarimetric rendering
and real captures. 
Analytic models provide diversity, but their hand-crafted polarization
behavior can deviate from real measurements, especially when dielectric~\cite{baek2018simultaneous} and
metallic~\cite{walter2007microfacet} circular-polarization responses are modeled separately. 
By expanding measured pBRDFs, our dataset better preserves realistic
polarimetric reflectance behavior. 
As shown in Fig.~\ref{fig:ablation}, the model trained on analytic
pBRDFs retains
illumination-dependent artifacts in albedo and roughness, whereas our
measured-pBRDF-based expansion yields cleaner material estimates and more
consistent reflectance.

\begin{table}[t]
\centering
\caption{\textbf{Ablation study.} 
The upper section ablates input encoding components; the lower section compares pBRDF training sources, averaged over two held-out test splits.}
\label{tab:ablation}
\resizebox{0.85\linewidth}{!}{
    \begin{tabular}{l|cccc}
    \toprule
     & Albedo & Roughness & Metallicity & Normal \\
    \cmidrule(lr){2-5}
     Variant & PSNR $\uparrow$
     & RMSE $\downarrow$
     & RMSE $\downarrow$
     & MAE $\downarrow$ \\
    \midrule
    Uniform white illumination     & 21.36 & 0.1166 & 0.3218 & 20.08 \\
    + Spectral multiplexing        & 21.61 & 0.1024 & 0.3010 & 12.45 \\
    + Polarization decomposition   & 22.22 & 0.0889 & 0.2889 & 11.97 \\
    + CP (ours)                    & \textbf{23.90} & \textbf{0.0781} & \textbf{0.2395} & \textbf{11.45} \\
    \midrule
    Analytic pBRDF                 & 17.14 & 0.2044 & 0.2885 & \textbf{14.81} \\
    Expanded pBRDF (ours)          & \textbf{18.06} & \textbf{0.1520} & \textbf{0.2836} & 18.14 \\
    \bottomrule
    \end{tabular}
}
\end{table}

\subsection{Real-world Reconstruction and Relighting}
We further assess generalization beyond the display illumination. 
Figs.~\ref{fig:rotating_light} and~\ref{fig:env_map} show relighting under a rotating point light and environment maps, respectively, where our predictions preserve material identity and coherent highlight behavior under diverse incident illumination. 
Fig.~\ref{fig:face} further demonstrates our snapshot capability on a dynamic face sequence: because our method requires only a single snapshot per frame, it can be applied independently to each frame without temporal alignment or registration, remaining robust to temporal variation and non-rigid deformation.

\section{Conclusion}
\label{sec:conclusion}
We presented a polarimetric display-based inverse rendering framework that estimates geometry and physically based material parameters from a single-shot spectro-polarimetric observation. 
Our polarization-aware decomposition reduces reflectance ambiguity by providing complementary unpolarized, linear-polarized, and circular-polarized cues.
Our approach performs robustly across synthetic and real data and generalizes to dynamic scenes with per-frame estimation.
In addition, the proposed pBRDF expansion framework provides diverse and physically valid training data for learning-based inverse rendering.
Overall, our results highlight the potential of polarization-aware spectral multiplexing as a scalable and practical solution for inverse rendering beyond controlled multi-shot capture setups.

\paragraph{Limitations}
Our method has several limitations. 
First, it relies on accurate polarization cues from the display--camera system and assumes limited spectral and polarization cross-talk. 
Non-ideal optical behavior, such as imperfect LCD polarization, inter-channel leakage, sensor noise, or ambient illumination, can degrade the recovered signals and estimation accuracy.
Second, polarization cues may be insufficient for materials with weak or ambiguous responses, such as mirror-like surfaces or complex layered materials.
Third, although PCA-based pBRDF expansion increases training diversity, it remains bounded by the coverage of the measured dataset; materials with strong anisotropy, subsurface scattering, or multi-layered structure may therefore lead to reduced generalization.
Finally, while our method supports per-frame dynamic estimation, severe motion blur, rapid non-rigid deformation, and heavy occlusions can degrade performance.

\clearpage


\clearpage
\newpage

\begin{figure*}
  \vfill
    \begin{minipage}{0.49\textwidth}
        \vspace{-1mm}
        \includegraphics[width=\columnwidth]{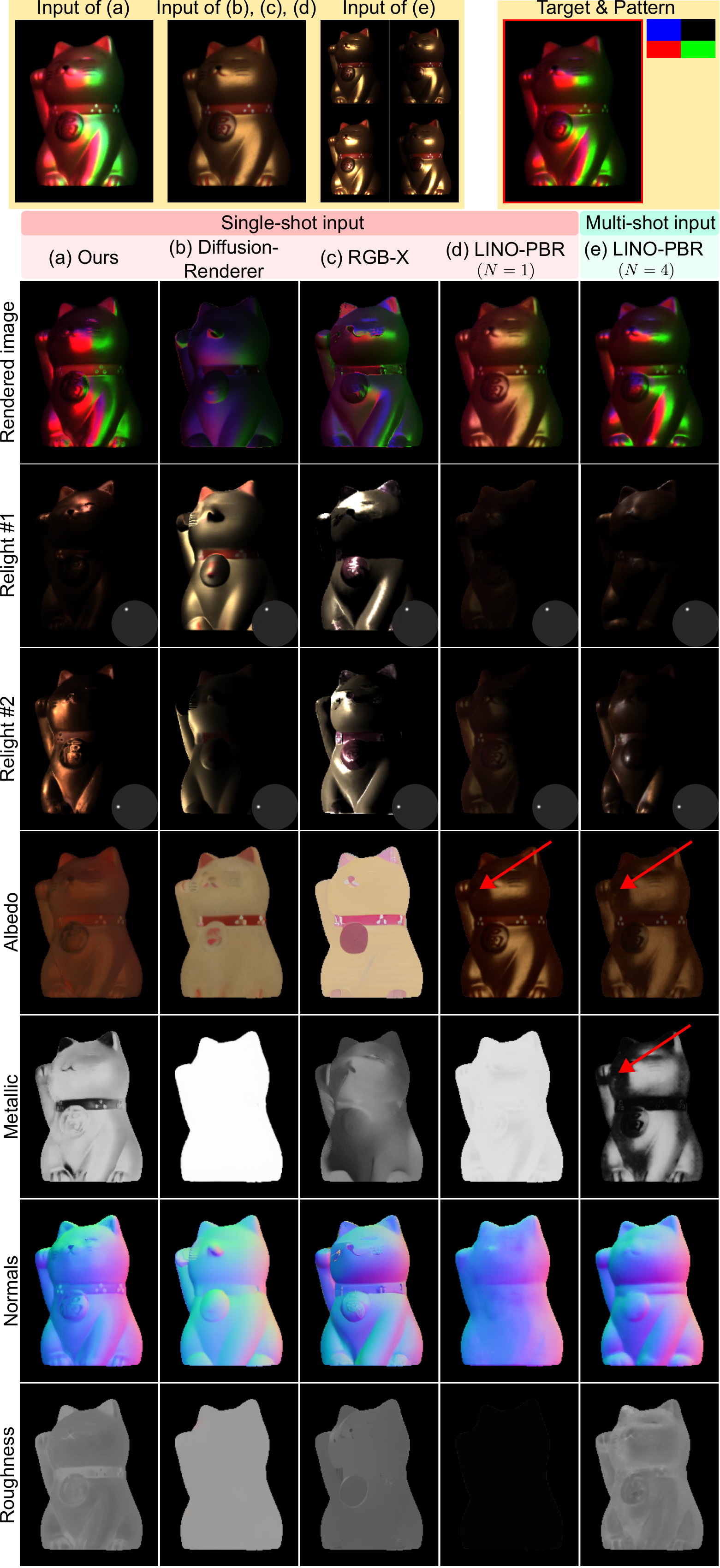}
        \vspace{-7mm}
        \caption{\textbf{Comparison on a real scene.}
        We show estimated PBR maps and renderings under both a known display pattern and an out-of-distribution point light.
        For metallic surfaces at grazing angles, diffuse reflection is largely attenuated, causing methods (d) and (e) to bake the observation into their albedo and metallicity estimates. 
        In contrast, our method correctly separates the materials of the ear and necklace and produces consistent reflectance estimates, leading to rendered images closest to the target and better preserving metallic reflectance at grazing angles.}
        \label{fig:comparison}
    \end{minipage}
    \hspace{1.5mm}
    \begin{minipage}{0.49\textwidth}
        \includegraphics[width=\columnwidth]{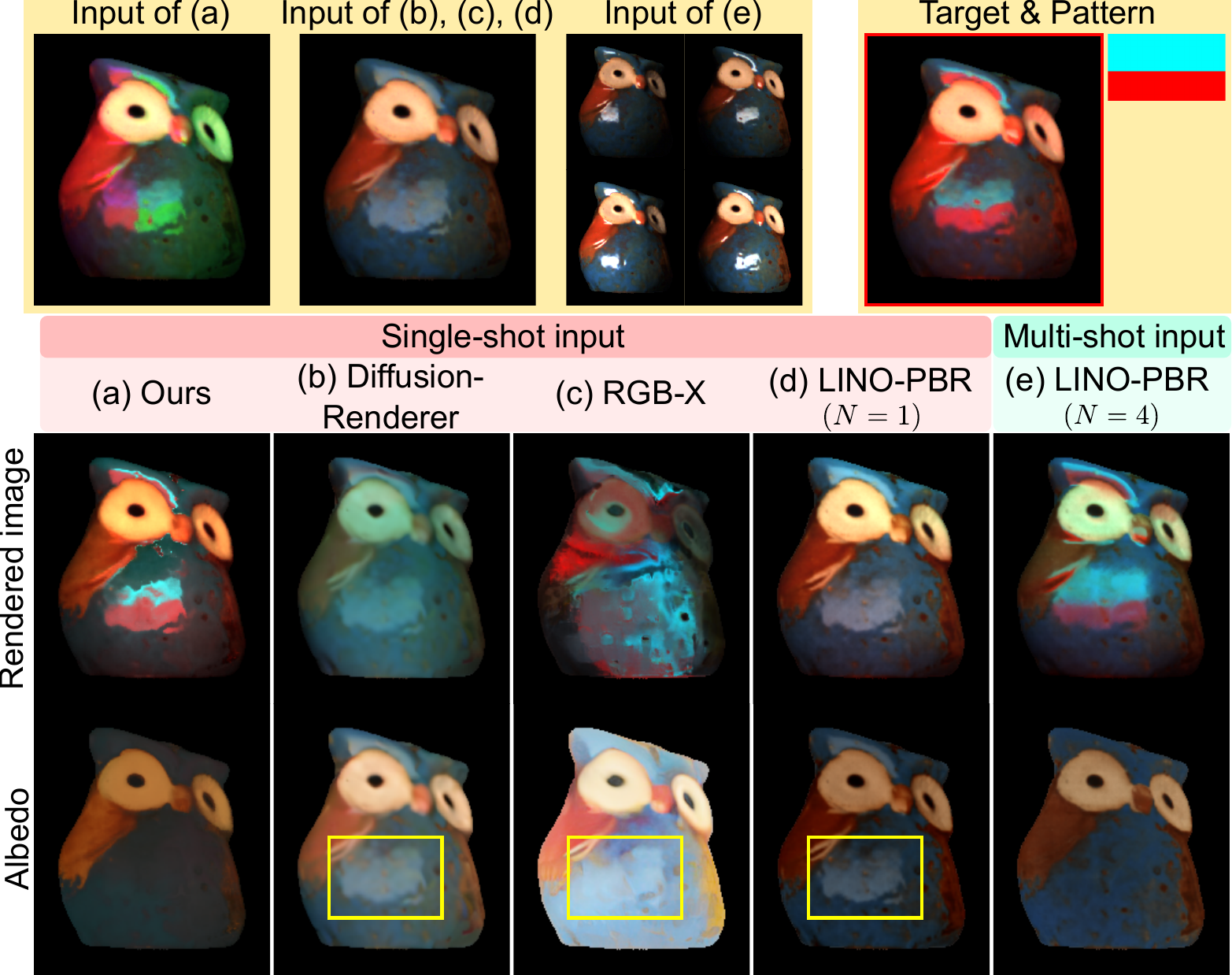}
        \vspace{-7mm}
        \caption{\textbf{Diffuse--specular disambiguation.}
        Single-image configurations (b), (c), (d) cannot observe lighting-dependent specular variations and therefore tend to entangle them with diffuse reflectance. This ambiguity propagates to the overall parameter estimation, producing rendered images that deviate noticeably from the target. In contrast, our method embeds lighting variation into the spectral channels of a single image through an RGB binary pattern, thereby alleviating this ambiguity.
        }
        \label{fig:albedo}
    
        \vspace{2mm}
        \includegraphics[width=\columnwidth]{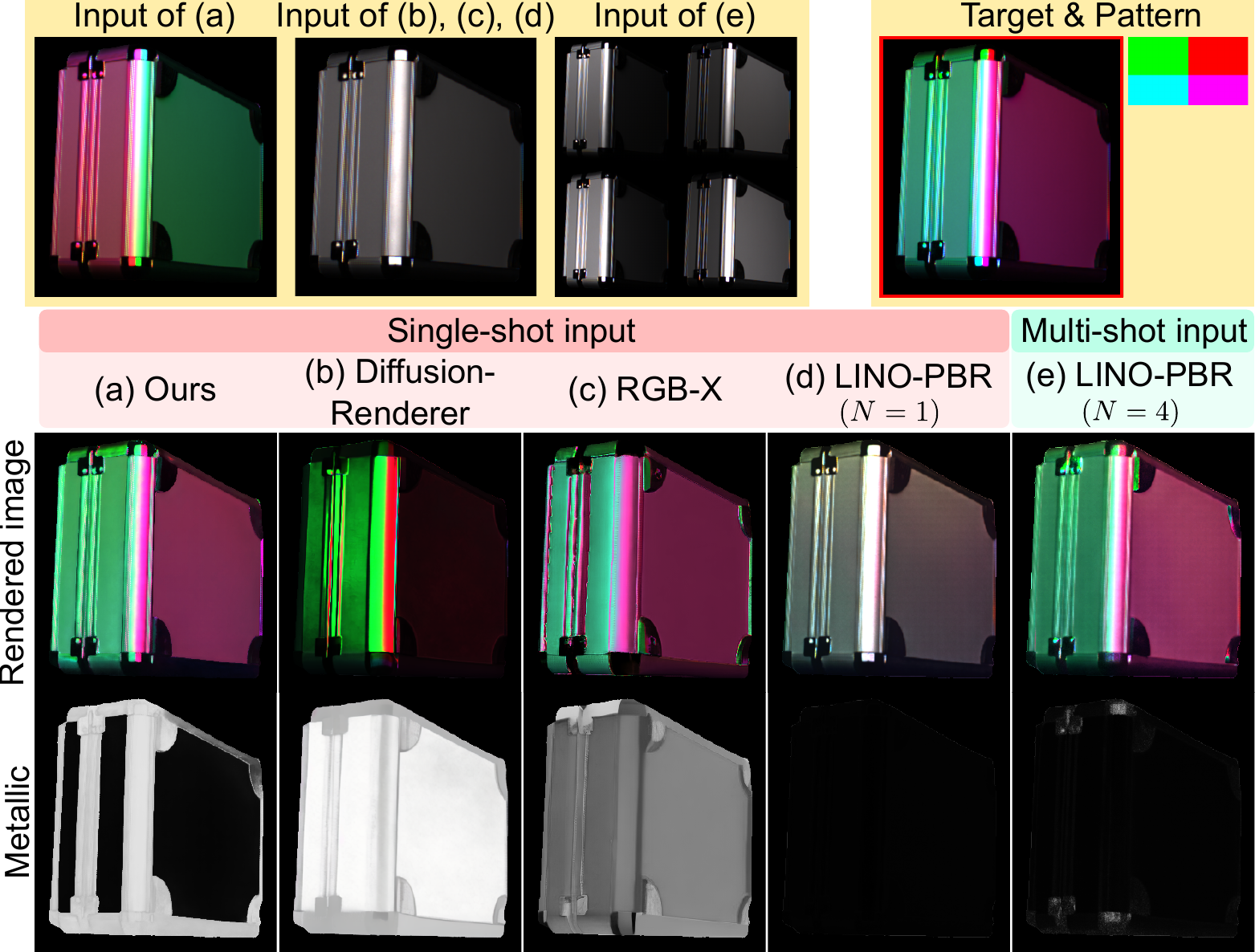}
        \vspace{-6mm}
        \caption{\textbf{Metallic disambiguation.}
        Methods that do not exploit CP cues struggle to distinguish the metallic parts of the object.}
        \label{fig:metallic}

        \includegraphics[width=\columnwidth]{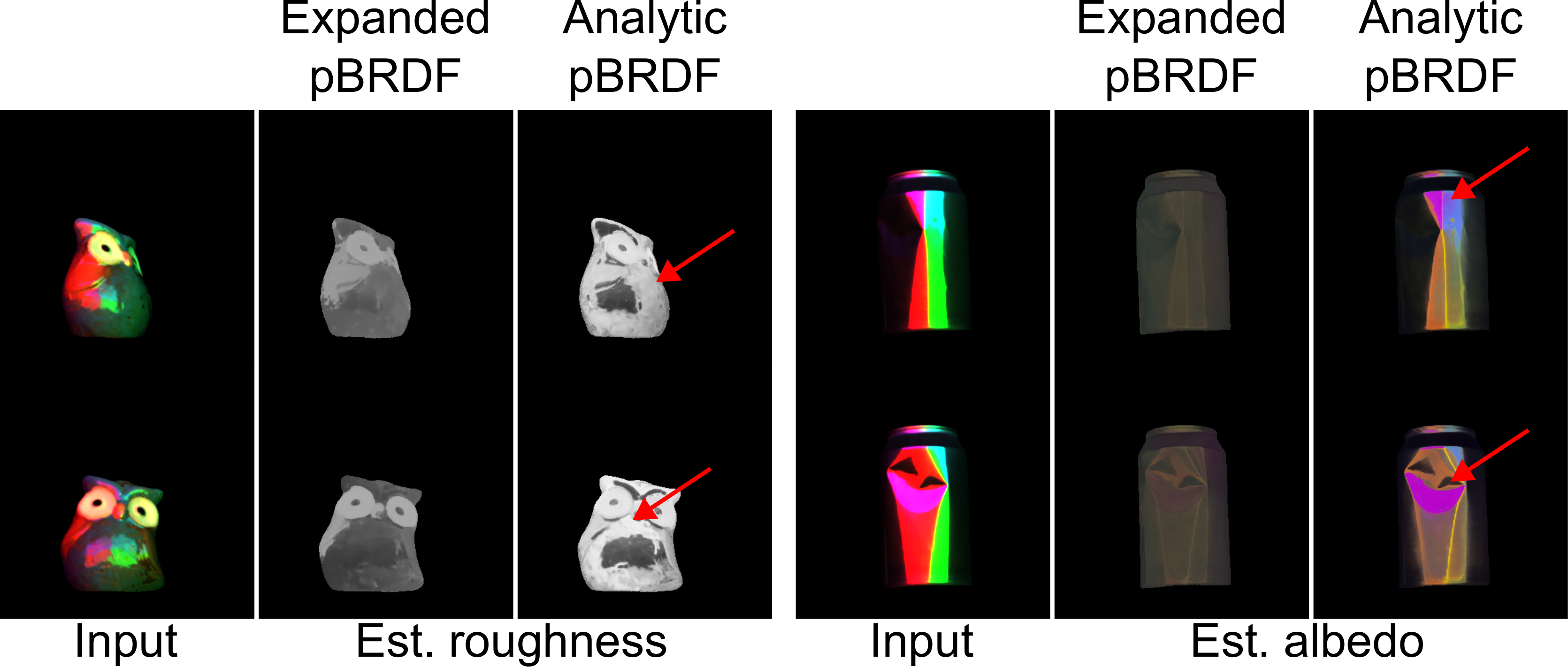}
        \vspace{-8mm}
        \caption{\textbf{Ablation on the expanded pBRDF dataset.}
        Unlike our measured-pBRDF-based expansion, training on analytic pBRDFs fail to faithfully disentangle real--world light--material interactions.}
        \label{fig:ablation}
    \end{minipage}

\end{figure*}

\clearpage
\newpage

\begin{figure*}[t]
    \begin{minipage}[t]{0.49\textwidth}
        \centering
        \includegraphics[width=\columnwidth]{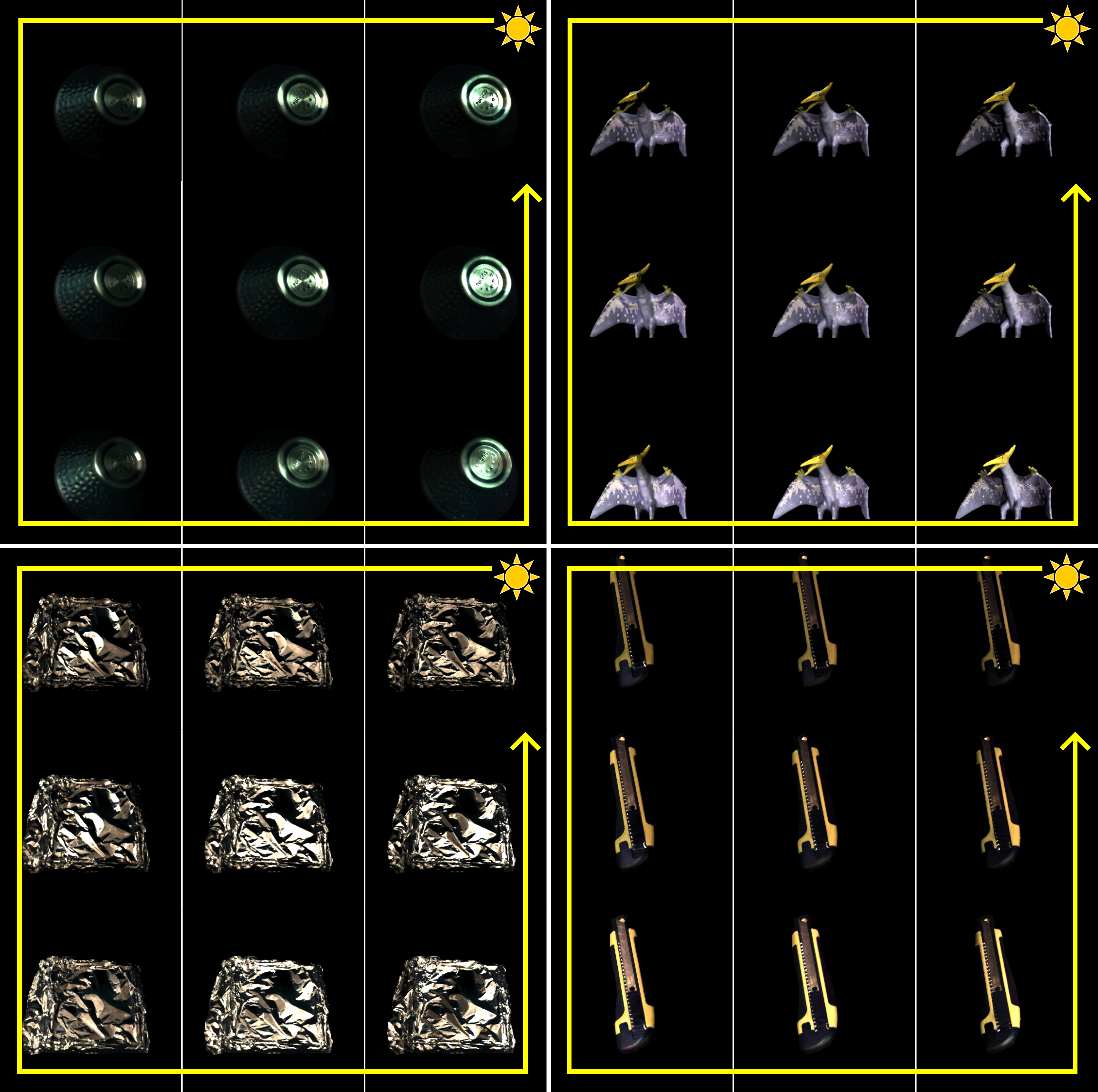}
        \vspace{-4mm}
        \caption{
        \textbf{Relighting under a rotating point light.}
        We relight the recovered PBR parameters of a real scene with a rotating point light.
        Highlights smoothly follow the illumination direction, while material appearance remains stable.
        }
        \label{fig:rotating_light}
        \vspace{1mm}
        
    \end{minipage}
    \hspace{1.5mm}
    \begin{minipage}[t]{0.49\textwidth}
        \centering
        \includegraphics[width=\textwidth]{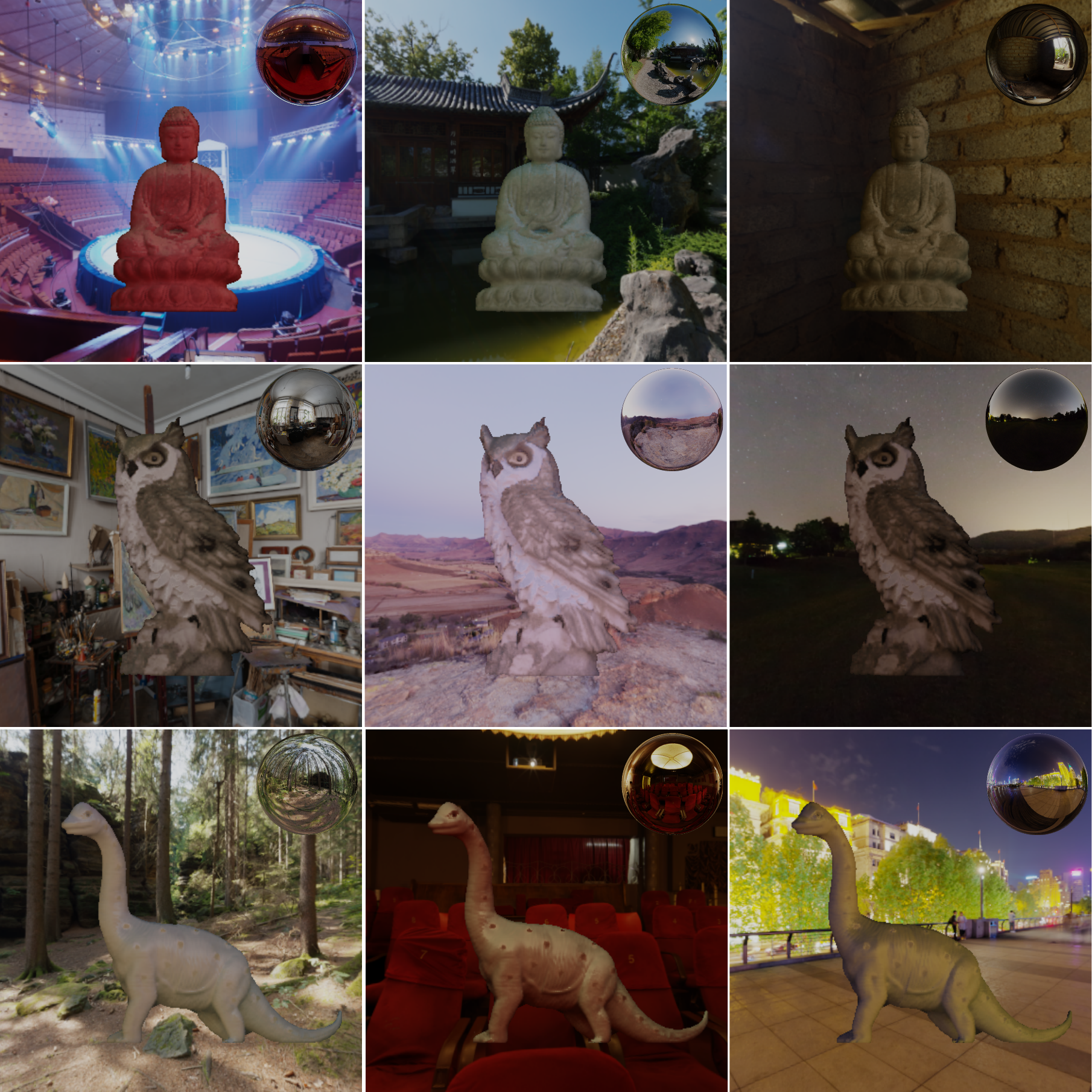}
        \vspace{-7mm}
        \caption{
        \textbf{Relighting under environment maps.}
        We render the recovered PBR maps of real captured objects under several environment maps that differ substantially from the display illumination used during capture.
        Our method preserves consistent material identity and produces coherently oriented highlights, yielding relighting results that blend naturally with the composited backgrounds without method-specific post-processing.
        }
        \label{fig:env_map}
    \end{minipage}
\end{figure*}

\begin{figure*}[b]
\centering
\includegraphics[width=\textwidth]{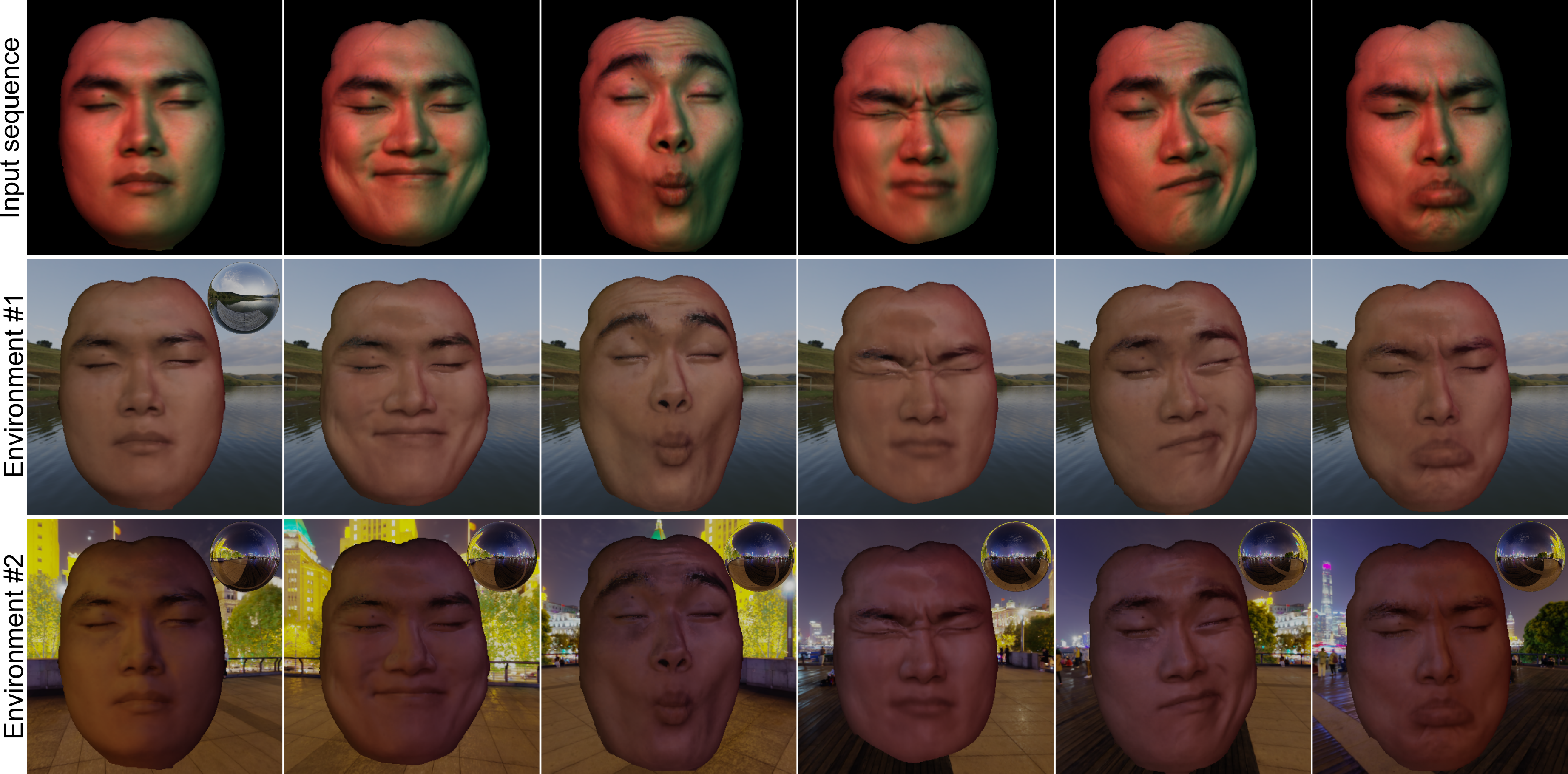}
\vspace{-6mm}
\caption{
\textbf{Relighting a dynamic face sequence.}
Our snapshot acquisition enables frame-wise relighting of dynamic scenes,
where multi-frame capture would require temporal alignment or registration
under facial motion. We apply our feed-forward pipeline independently to
each frame of a non-rigid face sequence and relight the predictions under
environment illumination. The results maintain consistent reflectance across
motion and varying illumination.
}
\vspace{-4mm}
\label{fig:face}
\end{figure*}
\clearpage

\bibliographystyle{plainnat}
\bibliography{references}
\clearpage

\end{document}